\documentstyle[twocolumn,prl,aps]{revtex}
\tighten
\begin{document}
\title{Adsorption Isotherms of Hydrogen: The Role of Thermal Fluctuations}
\author{Jens Vorberg, Stephan Herminghaus}
\address{Dept. of Applied Physics, University of Ulm, D-89069 Ulm,
Germany}
\author{Klaus Mecke}
\address{Fachbereich Physik, Universit\"at Gesamthochschule Wuppertal,
D-42097 Wuppertal, Germany}
\date{\today}
\maketitle
\vspace{1cm}

\begin{abstract}

It is shown that experimentally obtained isotherms of adsorption
on solid substrates may be completely reconciled with Lifshitz
theory when thermal fluctuations are taken into account. This is
achieved within the framework of a solid-on-solid model which is
solved numerically. Analysis of the fluctuation contributions
observed for hydrogen adsorption onto gold substrates allows to
determine the surface tension of the free hydrogen film as a
function of film thickness. It is found to decrease sharply for
film thicknesses below seven atomic layers.
\end{abstract}

\vspace{0.5cm}

Solid surfaces exposed to a gaseous environment attract the gas
molecules by virtue of molecular forces, giving rise to a thin
film of adsorbed molecules under almost any condition. The most
important attractive force involved is the van der Waals force,
which is due to the quantum mechanical zero point fluctuations of
the electron shells of the molecules. It is present for all
materials and extends over distances much larger than a molecular
diameter. It was thus tacitly accepted for a long time that the
theory of Frenkel, Halsey and Hill (FHH)
\cite{frenkel,halsey,hill}, which is based on the van der Waals
force alone, accurately describes adsorption isotherms, i.e., the
dependence of the thickness of the adsorbed film as a function of
the partial pressure of the adsorbed species in the surrounding
medium. It predicts that the thickness, $d$, of the adsorbed film
varies as
\begin{equation}
d = (\frac{\alpha}{k_BT \ln \frac{p_0}{p}})^{1/3}
\label{FHH}
\end{equation}
where $\alpha$ is the van der Waals constant of the system, $p$ is the gas pressure, and $p_0$ is the saturated
vapor pressure of the film material. The power $1/3$ on the r.h.s. of eq.~\ref{FHH} stems from the fact that according to the Lifshitz theory of dispersion forces, the adsorption energy should scale as the inverse third power of the film thickness \cite{lifshitz,dlp}. This holds for the film thicknesses to be discussed here, for which retardation effects can be safely neglected \cite{dlp,sabiskyanderson}. Deviations from FHH theory are to be expected only when the temperature is close to the critical point of the adsorbing substance, such that the width of the free film surface may be comparable to the film thickness.

On the contrary, experimentally determined adsorption isotherm revealed considerable deviations from FHH theory for virtually all systems studied \cite{lando,christenson,bradberry,panellakrim}. Although it has been mentioned before that
thermal fluctuations may be the cause for the observed discrepancies
\cite{panellakrim,meckekrim}, this possibility has not yet been
studied in detail experimentally. It is the purpose of the present paper to show that thermal fluctuations of the free film surface are very probably the cause of the above mentioned discrepancies. If these fluctuations are taken into account, the adsorption isotherms may be completely reconciled with theory.

As a model system, we have chosen hydrogen films adsorbed on gold
substrates. Films have been prepared in a helium flow cryostat at
temperatures around the triple point, ranging from 10 to 14 K
(T$_3$ = 13.952 K). The thickness of the adsorbed films has been
determined via surface plasmon resonance spectroscopy, as it was
done successfully before \cite{annalen}. The substrates consisted
of approximately 50 nm thick gold films prepared by evaporation
onto polished glass. The adsorbed hydrogen shifts the surface
plasmon resonance of the metal surface, which allows to determine
the coverage, i.e., amount of material adsorbed per unit area
\cite{annalen,raether}. This can be converted into an effective
film thickness assuming the film to have bulk material density.

Inspection of the gold substrates by scanning tunnelling
microscopy revealed a roughness and a lateral correlation length
similar to what was found before by other authors
\cite{gold1,gold2}. The surfaces exhibit large crystal facets,
such that at temperatures below the roughening transition of
hydrogen ($\approx$ 9 K \cite{tobepublished}), two dimensional
phase transitions in adsorbed single layers of hydrogen
\cite{tobepublished,hess} could be easily observed. From the
substrate topography, it could be estimated by the Kelvin equation
\cite{kelvin} that measurements of coverage might lose reliability
at thicknesses well above 5 nm due to capillary condensation
effects. Consequently, only films with a thickness up to 5 nm were
used for analysis. As it will be seen below, effects of
fluctuations are most important at much smaller thickness, where
roughness effects can be safely neglected.

To start with, we mention that both above and below the triple
point temperature, $T_3$, the thick hydrogen film which adsorbs
close to saturation is in the liquid phase
\cite{annalen,rieutord}. This is seen qualitatively from the fact
that if one takes for $p_0(T)$ in eq.~\ref{FHH} the saturated
vapor pressure of the {\it liquid} phase, which must below $T_3$
be obtained by extrapolation of the liquid/vapor coexistence line,
all isotherms fall onto a single master curve. This is
demonstrated in fig.~\ref{mastercurve}, which shows the data of
ten isotherms obtained for temperatures ranging from 10.190 K to
14.067 K, plotted as a function of reduced pressure, $p/p_0(T)$.
Since the saturated vapor pressure of the solid phase is always
below that of the liquid phase, the divergence of the isotherm is
not reached for temperatures below $T_3$, such that the film
thickness at solid/vapor coexistence is finite. This thickness at
saturation increases with increasing temperature, and diverges for
$T \rightarrow T_3$. This is triple point wetting \cite{annalen}.

As one can see in fig.~\ref{mastercurve}, the mutual agreement of
the isotherms is very good for sufficiently thick films. The van
der Waals constant is found to be $7 \pm 3$ $\rm K(nm)^3$, which
is consistent with the theoretical estimate of 8 $\rm K(nm)^3$
\cite{chengcole}. However, closer inspection of the data reveals
systematic discrepancies which are particularly obvious at
thicknesses below 2 nm, as it can be seen from the inset. This is
revealed more clearly in fig.~\ref{ourmodelfit}, where the data of
a single isotherm (T = 13.964 K) are plotted such that the
behaviour described by eq.~\ref{FHH} would yield a straight line.
The characteristic negative curvature of the isotherm in this plot
demonstrates the deviation, and is similar to what has been
described before by other authors \cite{christenson,panellakrim}.

The data are very well represented by the solid curve, which is
obtained from a refined model taking into account thermal
fluctuations in film thickness \cite{meckekrim}. The statistical
approach is based on a SOS-model which has extensively been
applied to multilayer-stepped adsorption isotherms
\cite{weeks,weeks2,weeks3}. We assume the substrate to consist of
a square lattice of $N$ adsorption sites  $i=1,\ldots, N$ and
thickness $d_i=a_0 n_i$ of the adsorbed film at the site $i$
allowing $n_i\geq 0$ to be any non-negative integer where $a_0$
denotes the monolayer thickness. The restriction of gas molecules
to certain lattice sites is well proven for a monolayer and seems
to be justified for films only few layers thick, i.e. the regime
which we  are focussing on. The vapor is considered  to be a
homogeneous reservoir of molecules with chemical potential
$\mu=k_BT \log (p_0/p)$ and the adsorbed molecules are assumed  to
pile up at each site in columns, without forming overhangs or
vapor bubbles, which is reasonable for thin films and temperatures
well below the critical point. The statistics of the film
thickness  is then given by the partition sum $Z=\sum_{\{n_i\}}
\exp\left[-\beta {\cal H}(\{d_i\})\right]$, where the sum runs
over all configurations $n_i$ ($i$=1,...,$N$) of the film. The
Hamiltonian reads \cite{meckekrim}
\begin{equation}
{\cal H}(\{d_i\}) = \sum_i^N \left(d_i k_BT \log {p_0\over p} -
\sum_{\nu=1}^{n_i}{\alpha
\over z_\nu^3}  \right) + {\gamma \over 2}
\sum_{<ij>}(d_i-d_j)^2
\label{hamiltonian}
\end{equation}
where $z_\nu$ is the distance of the $\nu$-th layer to the
substrate. The surface tension $\gamma$ of the film-vapor interface
takes into account the  molecular interactions within the film, where
sum runs over nearest-neighbor
sites $<ij>$ only.

Assuming that fluctuations in the  film thickness  are not
relevant, one may minimize the energy  (\ref{hamiltonian})
yielding the most probable thickness $d$ given by the
FHH-isotherm, Eq. (\ref{FHH}). But the film-vapor interface is
always undulated due to thermal fluctuations which become
important for thin films where fluctuations are hindered by the
substrate. Thus, in order to perform the partition sum   we apply
a mean-field approximation, replacing $d_j$ in Eq.
(\ref{hamiltonian}) with its mean value $\bar{d}$. One obtains a
self-consistent equation for the mean thickness, which can be
solved by standard numerical procedures and can be applied to the
experimental data in a straightforward manner (see
Ffig.~\ref{ourmodelfit}). The solution is the adsorption isotherm
$\bar{d}(\mu;\alpha,a_0,\gamma)$ depending on   the Hamaker
constant $\alpha$, the monolayer thickness $a_0$, and the surface
tension $\gamma$. Of course, $\gamma(\bar{d})$ depends itself on
the  mean film thickness
 and may be determined experimentally when vapor pressure $p$ and  thickness
$\bar{d}$  is measured.

This SOS-model reproduces the results of Brunnauer, Emmett, and
Teller \cite{BET} as well as the FHH model in the  monolayer and
thick-film regimes, respectively, but for intermediate coverages,
a qualitatively different behavior is found which is governed by
thermal fluctuations of the film thickness, and is therefore
determined by the surface tension, $\gamma$. As one can see in
fig.~\ref{ourmodelfit}, the substrate-induced hindrance of such
fluctuations increases significantly the mean thickness $\bar{d}$
as compared to the most probable thickness $d$ given by the
FHH-isotherm. Most importantly, the isotherms obtained
experimentally are well described taking fluctuations into
account, without any further assumptions. Qualitatively,  the
effect of the fluctuations, i.e., the deviation from the FHH
isotherm, increases with increasing temperature, as expected. For
illustration, we have included in fig.~\ref{ourmodelfit} what is
obtained from our model if fluctuations are absent (dashed line).
This was done by setting $\gamma \rightarrow \infty$ in the
calculations. The impact of the fluctuations becomes particularly
obvious this way.

Let us now go a step further and use the fluctuation contributions
identified above to determine the surface tension, $\gamma$, of
the hydrogen/vapor interface. This turns out to be mathematically
difficult on the basis of eq.~\ref{hamiltonian}. Therefore, we
used a somewhat simpler approach based on the
fluctuation-dissipation theorem, which directly yields a more
convenient expression for the excess chemical potential due to
capillary waves \cite{kralchevsky}, and also contains $\gamma$ as
a parameter. It is less accurate than eq.~\ref{hamiltonian}
because it neglects the geometric limitation of the fluctuation
amplitude by the finiteness of the film thickness, but it suffices
for a rough evaluation of $\gamma$.

The result is shown in fig.~\ref{surfacetension}. While one would
naively expect  $\gamma$ to be constant, we see that it is
dramatically reduced for small thickness. The jump at 0.7 nm may
well indicate the boundary between the solidified part of the film
close to the wall \cite{annalen,rieutord} and the liquid. At
higher film thickness, a strong increase of the surface tension is
seen, which approaches the bulk value above 25 Angstroms,
corresponding to about 6.5 molecular layers (one layer is 3.8
Angstroms thick). We tend to be cautious as to the actual amount
by which the surface tension of the thin films is reduced, because
the method used for its determination is not as accurate as
eq.~\ref{hamiltonian}, as mentioned above. However, there is no
doubt according to our qualitative result that the surface tension
is considerably less for the thin film than for the bulk. It is
desirable to develop an analysis based on eq.~\ref{hamiltonian} to
check the result displayed in fig.~\ref{surfacetension}.

In conclusion, we have demonstrated, with hydrogen adsorbed on
gold as a model system, that adsorption isotherms may be
completely reconciled with Lifshitz theory if thermal fluctuations
of the free film surface are taken into account. This seems to
resolve a long standing discussion in the field of wetting forces,
and also provides a means of determining the surface tension of
molecularly thin liquid films. For hydrogen, we found a
substantial reduction of the surface tension with respect to the
bulk value. The exact physical nature of this effect must be left
to further studies.

This work was supported by the Deutsche Forschungsgemeinschaft within the priority program `Wetting and Structure Formation at Interfaces' and the Sonderforschungsbereich 262.

\vspace{1cm}

\begin{figure}[h]
\caption{Assuming the adsorbed film to be liquid, it is natural to use
the saturated vapor pressure of the liquid phase in the FHH
equation. In fact, this makes all isotherms fall onto a single master
curve, independent of temperature. The slight deviations in the flat
parts of the curve (around $p = 0.5 p_0$) are significant and may be
well explained by thermal fluctuations. They are shown clearly in the inset, which displays isotherms for 10.190 K (squares), 13.939 K (triangles), and 13.964 K (circles).}
\label{mastercurve}
\end{figure}

\begin{figure}[h]
\caption{A typical isotherm, obtained at T = 13.964 K, plotted such that an FHH isotherm according to eq.~\ref{FHH} would yield a straight line. The negative curvature is clearly visible. The solid curve is a fit of our model. For the dashed curve, the surface tension has been set to infinity in the model, in order to suppress fluctuation effects. The difference between the data and the dashed curve thus demonstrates the significant impact of fluctuations.}
\label{ourmodelfit}
\end{figure}

\begin{figure}[h]
\caption{The surface tension of hydrogen as a function of film
thickness, as obtained from the analysis of the fluctuation
contribution to the equilibrium film thickness. At small thickness, the surface tension is dramatically reduced, and approaches the bulk value only above 2.5
nm (about 6.5 ML).}
\label{surfacetension}
\end{figure}

\end{document}